\documentclass[lettersize,journal,]{IEEEtran}
\usepackage{amsmath,amsfonts}
\usepackage{algorithmic}
\usepackage{algorithm}
\usepackage{array}
\usepackage[caption=false,font=normalsize,labelfont=sf,textfont=sf]{subfig}
\usepackage{textcomp}
\usepackage{stfloats}
\usepackage{url}
\usepackage{verbatim}
\usepackage{graphicx}
\usepackage{cite}
\usepackage{booktabs}
\usepackage{tabularx}
\usepackage{hyperref}

\usepackage[capitalize,noabbrev]{cleveref}
\usepackage{longtable} 
\usepackage{multicol}
\usepackage{multirow}
\usepackage{enumitem}

\usepackage{xcolor}
\usepackage{tcolorbox}
\tcbuselibrary{skins, breakable}

\newtcolorbox{promptbox}[1][]{
    colback=gray!10, 
    colframe=gray!50, 
    arc=2pt, 
    boxrule=1pt, 
    fontupper=\footnotesize\ttfamily,  
    boxsep=2pt,                 
    left=3pt,                   
    right=3pt,                  
    top=2pt,                    
    bottom=2pt,                 
    title=\textbf{REPLACE ME}, 
    fonttitle=\sffamily\footnotesize,
    colbacktitle=black!70,
    enhanced,
    breakable,
    before upper={\parindent0pt}, 
    #1 
}

\usepackage{fvextra}

\begin{document}

\title{VeriInteresting: An Empirical Study of Model–Prompt Interactions in Verilog Code Generation}

\author{Luca Collini~\IEEEmembership{Graduate Student Member,~IEEE,}, Andrew Hennessee~\IEEEmembership{Graduate Student Member,~IEEE,} Patrick Yubeaton~\IEEEmembership{Graduate Student Member,~IEEE,} Siddharth Garg~\IEEEmembership{Member,~IEEE,} Ramesh Karri~\IEEEmembership{Fellow,~IEEE}

\thanks{Manuscript received April 19, 2021; revised August 16, 2021.}}

\markboth{Journal of \LaTeX\ Class Files,~Vol.~14, No.~8, August~2021}%
{Shell \MakeLowercase{\textit{et al.}}: A Sample Article Using IEEEtran.cls for IEEE Journals}


\maketitle

\begin{abstract}
Rapid advances in language models (LMs) have created new opportunities for automated code generation while complicating trade-offs between model characteristics and prompt design choices. In this work, we provide an empirical map of recent trends in LMs for Verilog code generation, focusing on interactions among model reasoning, specialization, and prompt engineering strategies. We evaluate a diverse set of small and large LMs, including general-purpose, reasoning, and domain-specific variants. Our experiments use a controlled factorial design spanning benchmark prompts, structured outputs, prompt rewriting, chain-of-thought reasoning, in-context learning, and evolutionary prompt optimization via Genetic-Pareto. Across two Verilog benchmarks, we identify patterns in how model classes respond to structured prompts and optimization, and we document which trends generalize across LMs and benchmarks versus those that are specific to particular model–prompt combinations.
\end{abstract}

\begin{IEEEkeywords}
Large Language Models, Verilog Generation, GEPA
\end{IEEEkeywords}

\section{Introduction}




The landscape of language models (LMs) is advancing rapidly, with new architectures, training paradigms, and inference-time techniques emerging continuously. In software engineering, LMs are dominant productivity tools, substantially improving code generation and iteration. This growth in software capability increases the demand placed on the hardware ecosystem that supports it.
Despite this demand, hardware engineers have not seen comparable productivity gains. Hardware design is a constrained, human-driven process, governed by the strict semantics of hardware description languages (HDLs) and complex power, performance, area, and cost trade-offs. Unlike Python, Verilog is verbose and structural, requiring models to maintain long-context coherence for precise signal connectivity. The stakes in hardware are higher; a plausible-looking Verilog module may simulate correctly in isolation, yet violate timing constraints or introduce race conditions that surface in late-stage verification or fabrication, raising reliability concerns.

Applying LMs to hardware design is complicated by a scarcity of structured data. Most hardware designs constitute high-value, closed-source intellectual property, limiting open-source training data and restricting retraining or proprietary fine-tuning. Many organizations are reluctant to rely on third-party models or cloud-based services, necessitating inference-time methods that extract maximal performance from existing LMs without access to private data.

In this work, we present a large-scale empirical study of LMs for Verilog generation, focusing on the interplay of reasoning capabilities, prompt design, and prompt optimization. We evaluate 18 LMs across two Verilog benchmarks with complementary evaluation protocols (simulation-based testing and formal equivalence checking): 7 commercial large language models (LLMs) and 11 open-source small language models (SLMs), including 4 SLMs fine-tuned for Verilog. Our study spans prompting strategies such as structured prompting~\cite{khattab2024dspy}, chain-of-thought reasoning~\cite{wei2023chainofthought}, in-context learning~\cite{brown2020language}, and prompt refinement~\cite{acecoder, srivastava-etal-2024-instances}. Finally, we use Genetic-Pareto (GEPA) prompt optimization~\cite{agrawal2025gepareflectivepromptevolution}---an inference-only improvement method---to optimize system prompts without violating IP constraints or incurring fine-tuning costs.

To address this productivity gap and provide actionable deployment guidelines for Electronic Design Automation (EDA) teams, we formulate four research questions that isolate the effects of model scale, domain specialization, and inference-time prompting:
\begin{itemize}[leftmargin=*, topsep=2pt, itemsep=0pt]
  \item \textbf{RQ1:} How does model scale compare to domain-specific specialization under limited training data?
  \item \textbf{RQ2:} How sensitive is Verilog generation performance to prompting strategies?
  \item \textbf{RQ3:} Do finetuned LMs outperform strongly prompted general-purpose LMs?
  \item \textbf{RQ4:} How stable are trends across Verilog benchmarks?
\end{itemize}

Ultimately, our findings reveal that Verilog generation is a distinct task regime where software-centric prompt engineering often hits a fundamental ``capacity ceiling.'' We demonstrate that while inference-time optimization can narrow the gap for general-purpose models, highly specialized Verilog models can suffer from catastrophic forgetting when subjected to complex structural prompting. Based on these insights, we outline practical guidelines for deployment to help practitioners navigate the trade-offs between IP-secure local fine-tuning and cloud-based prompt optimization in modern RTL workflows.

\section{Related Works}
Language models have demonstrated strong baseline capabilities on programming and software engineering tasks, including code generation, completion, repair, and reasoning over complex specifications \cite{Jiang_2026, hasan2026}. 
 
Emerging research shows that prompt design and inference-time strategies can improve model performance on programming tasks. Many techniques have been explored, including structured prompts that enforce intermediate representations or schemas \cite{li2023SCoT}, retrieval-augmented prompting that incorporates external documentation \cite{zhou2023docprompt}, multi-stage pipelines that rewrite, refine, or debug generated programs \cite{chen2023selfdebug}, and model-aware in-context learning  \cite{li2023modelAwareICL}. More recently, automated prompt optimization methods have been proposed that iteratively refine prompts using task-level feedback to improve performance on code generation tasks~\cite{GEPAopenACC, prochemy}. These results suggest that performance on coding tasks is shaped not only by model capacity, but also by how effectively the task is specified and scaffolded at inference time.

In parallel, there has been growing interest in applying LMs to hardware design workflows, including HDL generation from natural language specifications~\cite{thakur2023verigen, liu2024rtlcoder, lu2023rtllm}, simulation- and formal-guided verification, and iterative debug loops~\cite{flag_}. While tasks such as Verilog generation are syntactically similar to conventional code generation, they are semantically distinct; hardware descriptions encode concurrency, timing, and microarchitectural behavior, and are evaluated through downstream synthesis and verification toolchains. As a result, hardware design imposes stricter correctness requirements, deeper dependence on external tools, and heightened sensitivity to subtle semantic errors that may not be apparent from code structure alone. These differences raise a natural question: to what extent do prompting and reasoning techniques that have proven effective in software-centric settings transfer to hardware design tasks?

To our knowledge, this question has not yet been systematically examined across models and benchmarks. We address this gap through a controlled, large-scale evaluation of prompting and inference-time optimization strategies for Verilog generation across multiple benchmarks and a diverse set of model classes, including commercial frontier LMs, open-source baselines, and Verilog-specialized LMs. Because these benchmarks differ in task formulation and evaluation protocols, our design enables us to (i) assess robustness of prompting gains across LMs and settings, and (ii) analyze how LM specialization and embedded reasoning influence the effectiveness of prompt engineering.

\section{Background}
\label{sec:background}

Generating Register-Transfer Level (RTL) code differs fundamentally from standard software code generation, as hardware correctness requires guarantees across all possible input states rather than validation over a finite test set. Accordingly, Verilog generation benchmarks vary not only in task sources and specification styles, but also in their evaluation protocols.
The most common approach is dynamic simulation, in which generated designs are compiled and executed against testbenches, and outputs are compared to a reference for a limited set of inputs. While analogous to software unit testing, simulation provides only partial coverage and may miss corner-case errors.
An alternative is formal equivalence checking, which uses symbolic methods to prove that a generated circuit is functionally identical to a trusted reference implementation for all input states. Although this offers a stronger notion of correctness, it relies on  a golden design and incurs a higher computational cost.

\subsection{Simulation-Based Evaluation: Verilog Eval}

Verilog Eval~\cite{liu2023VerilogEval} represents the standard for simulation-based evaluation. Originally released in 2023, it comprises 156 tasks from the HDLBits platform, ranging from basic logic gates to complex finite state machines. 
Verilog Eval uses the Icarus Verilog simulator to check for functional correctness. A generated design is deemed correct if its transient simulation outputs match those of a reference solution across a fixed set of test vectors. Recent updates in ``Revisiting Verilog Eval''~\cite{Pinckney25} have modernized this framework to support full specification-to-RTL generation (rather than just code completion) and added infrastructure to evaluate in-context learning strategies. 

\subsection{Verification-Based Evaluation: VeriThoughts}

To overcome the coverage limitations of simulation, VeriThoughts~\cite{yubeaton2025verithoughtsenablingautomatedverilog} introduces a benchmark grounded in formal verification. Unlike Verilog Eval, which checks input/output pairs, VeriThoughts uses Yosys~\cite{Wolf2013YosysAFV} open synthesis suite for logical equivalence checking (LEC). LEC verifies that the generated RTL is functionally equivalent to the golden RTL.


\subsection{LM Adaptation: Training to Inference}

Optimizing LLMs for Verilog generation generally follows two complementary strategies: training-time adaptation, which updates model parameters, and inference-time adaptation, which refines prompts without modifying the LM.

Training-time adaptation includes fine-tuned LMs such as VeriReason~\cite{wang2025verireasonreinforcementlearningtestbench} and VeriThoughts~\cite{yubeaton2025verithoughtsenablingautomatedverilog}, both introduced at NeurIPS in 2025 and representative of the current state of the art in domain-adapted LMs for RTL generation. VeriReason is optimized against simulation-based verification like Verilog Eval; it employs Guided Reward Proximal Optimization (GRPO)~\cite{shao2024deepseekmathpushinglimitsmathematical} to fine-tune base LMs (e.g., Qwen2.5-Coder) using testbench outputs as a reward signal, while also enforcing structural correctness through abstract syntax tree constraints. VeriThoughts models are fine-tuned on a dataset of formally verified specification-to-RTL pairs with accompanying reasoning traces.

Inference-time strategies optimize LM behavior without changing weights. Standard approaches include zero-shot and few-shot prompting~\cite{brown2020language}, as well as chain-of-thought~\cite{wei2023chainofthought} techniques. GEPA~\cite{agrawal2025gepareflectivepromptevolution} represents an advanced inference-time method, treating prompts as parameters and iteratively refining them via evolutionary search. 

\begin{table*}[thb]

\caption{Evaluated LMs and decoding parameters. For OpenAI models, decoding parameters are not disclosed and are not user-configurable for reasoning models.}
\label{tab:models}

\centering
\small
\begin{tabularx}{\linewidth}{l l X r r}

\specialrule{1.2pt}{0em}{0.4em}

\textsc{Model} & \textsc{Abbrev.} & \textsc{Description} & \textsc{Temp} & \textsc{Top P} \\

\specialrule{1.2pt}{0.2em}{0.4em}

GPT-4.1 nano & GPT-4.1n & General-purpose, not reasoning-specialized & 0.6 & 1.00 \\
GPT-4.1 mini & GPT-4.1m & General-purpose, not reasoning-specialized & 0.6 & 1.00 \\
GPT-5 nano & GPT-5n & Hybrid general-purpose + light embedded reasoning & \textemdash{} & \textemdash{} \\
GPT-5 mini & GPT-5m & Hybrid general-purpose + strong embedded reasoning & \textemdash{} & \textemdash{} \\
Gemini 3 Flash & Gem-Fl & Hybrid general-purpose + light embedded reasoning & 1.0 & 0.95 \\
Gemini 3 Pro & Gem-Pro & Hybrid general-purpose + strong embedded reasoning & 1.0 & 0.95 \\
Claude Sonnet 4 & Claude-S4 & Hybrid general-purpose + moderate embedded reasoning & 0.6 & 0.99 \\

\specialrule{0.2pt}{0.2em}{0.2em}

Qwen2.5-3B & Qw-3 & General-purpose model (small scale) & 0.6 & 0.80 \\
Qwen2.5-7B & Qw-7 & General-purpose model (mid scale) & 0.6 & 0.80 \\
Qwen2.5-14B & Qw-14 & General-purpose model (large scale) & 0.6 & 0.80 \\
Qwen2.5 Coder Instr.-3B & QwC-3 & Code-specialized, instruction-tuned model (small scale) & 0.6 & 0.80 \\
Qwen2.5 Coder Instr.-7B & QwC-7 & Code-specialized, instruction-tuned model (mid scale) & 0.6 & 0.80 \\
Qwen2.5 Coder Instr.-14B & QwC-14 & Code-specialized, instruction-tuned model (large scale) & 0.6 & 0.80 \\
DeepSeek CoderV2 Instr.-16B & DSC-16 & Code-specialized, instruction-tuned model (large scale) & 0.6 & 0.95 \\

VeriReason-3B & VR-3 & Verilog-specialized reasoning model (SFT + GRPO, small scale) & 0.6 & 1.00 \\
VeriReason-7B & VR-7 & Verilog-specialized reasoning model (SFT + GRPO, mid scale) & 0.6 & 1.00 \\
VeriThoughts-7B & VT-7 & Verilog-specialized reasoning model (SFT only, mid scale) & 0.6 & 0.80 \\
VeriThoughts-14B & VT-14 & Verilog-specialized reasoning model (SFT only, large scale) & 0.6 & 0.80 \\

\specialrule{1.2pt}{0.2em}{0.4em}

\end{tabularx}

\end{table*}









\section{Experimental Setup}






\subsection{Benchmarks}
We evaluate on two complementary specification-to-RTL benchmarks introduced in \cref{sec:background}: \textsc{Verilog Eval v2}, which uses simulation-based testbenches to check functional correctness, and \textsc{VeriThoughts}, which uses formal equivalence checking against a golden reference implementation. Using both benchmarks lets us distinguish improvements that benefit a specific model due to training biases and enables stronger claims about generalizability (RQ4).

\subsection{LMs and Adaptation Regimes}
\cref{tab:models} summarizes the 18 evaluated LMs, spanning: (i) commercial frontier LMs, (ii) open-source general-purpose and code-specialized LMs across multiple parameter scales, and (iii) Verilog-specialized LMs obtained via training-time adaptation (e.g., SFT and/or RL with feedback). This LM suite operationalizes comparisons between scale and specialization (RQ1), and between fine-tuned Verilog LMs and strongly prompted general-purpose baselines (RQ3).

\subsection{Prompting Axes and Experimental Conditions}
We study inference-time adaptation through a controlled set of prompting strategies. 
Each benchmark is evaluated under:
(i) a \textbf{baseline} single-pass prompt from the benchmark (Base);
(ii) \textbf{structured prompting} (Struct) that enforces a consistent task signature and modular prompt components;
(iii) \textbf{prompt refinement} (Refine) via a two-stage pipeline---inspired by~\cite{acecoder} and~\cite{srivastava-etal-2024-instances}---that first rewrites/clarifies the specification and then generates RTL;
(iv) optional \textbf{chain-of-thought} reasoning (CoT) inserted before code emission; and
(v) \textbf{in-context learning} (ICL) by adding exemplars to the prompt.

Our structured prompting and refinement pipelines are implemented using DSPy~\cite{khattab2024dspy}, which enforces structured I/O through explicit task signatures and modular predictors. This provides a programmatic mechanism for specifying intermediate representations and composing multi-stage generation, improving reproducibility and reducing prompt brittleness compared to ad-hoc template engineering. We include system prompts in~\Cref{app:sys_prompts}.

We apply CoT to both structured prompting and prompt refinement, yielding five prompting setups. Each setup is run with and without ICL, resulting in 10 experimental conditions per LM. This factorial design isolates sensitivity to prompt design (RQ2) while keeping the benchmark, model, and decoding configuration fixed.

ICL uses a fixed set of three exemplar prompt–response pairs, reused across all tasks and models (i.e., no per-task retrieval or selection). The exemplars were curated independently of the evaluation tasks and are not drawn from the \textsc{VeriThoughts} or \textsc{Verilog Eval v2} test sets, to avoid benchmark leakage. We selected exemplars to span common RTL motifs and interface styles, including (i) a clocked register interface with active-low asynchronous reset (Ex. 1), (ii) multi-output control logic with explicit reset behavior (Ex. 2), and (iii) a parameterized interface with synchronous signal registration (Ex. 3). This design aims to provide format/idiom anchoring while minimizing overlap with the benchmark task distributions.

Additionally, we apply GEPA to the structured prompt with CoT, which was the strongest strategy for both the top-performing SLM (Qwen2.5-14B) and the lowest-cost commercial model (GPT-4.1 Nano), to probe the best-case and cost-constrained regimes. This experiment evaluates how much inference-time optimization alone can improve low-cost models without fine-tuning. Prompts are optimized for the two target models and cross-evaluated to assess generalization. We use Gemini 3 Pro as the reflection model and sample 125 problems from the RTL Coder~\cite{liu2024rtlcoder} benchmark (100 for optimization, 25 for validation). GEPA is run in the ``heavy'' setting with a mini-batch size of five.

\subsection{Inference, Decoding, and Metric Evaluation}
For each model, we use the decoding hyperparameters listed in \cref{tab:models}. We selected a temperature of 0.6, which is common for code generation tasks~\cite{t06_1, t06_2}. We kept top p as the models' default. For commercial APIs where certain decoding controls are unavailable, we use the provider defaults and report them as not user-configurable. To estimate pass rates, we generate 20 candidates per task (via independent samples) for each experimental setup.
We report \emph{pass@$k$} (P@$k$) rates using the unbiased estimator proposed by~\cite{chen2021codex}. We use $k\in{1,5,10}$ to capture single-shot usability and benefits of lightweight resampling.

\subsection{Mapping Conditions to Research Questions}
Our setup is organized around four axes that instantiate the RQs: (RQ1) \textbf{scale vs. specialization} by comparing model families across sizes and Verilog-adapted variants; (RQ2) \textbf{prompt sensitivity} by sweeping Base/Struct/Refine and toggling CoT and ICL; (RQ3) \textbf{training-time vs. inference-time adaptation} by contrasting Verilog-fine-tuned models against general-purpose models under the strongest prompting conditions; and (RQ4) \textbf{benchmark stability} by repeating the full matrix across  benchmarks.

\section{Experimental Results}
\label{sec:results}

All numerical results are reported in \cref{app:num_results}. In this section, we focus on the qualitative trends and organize the discussion around our research questions (RQ1--RQ4). Our code and results are available in an \href{https://anonymous.4open.science/r/VeriInteresting-538D/}{anonymized repository}.

\subsection{RQ1: Scale vs. specialization}

\begin{figure}[h]
    \centering
    \includegraphics[width=\linewidth]{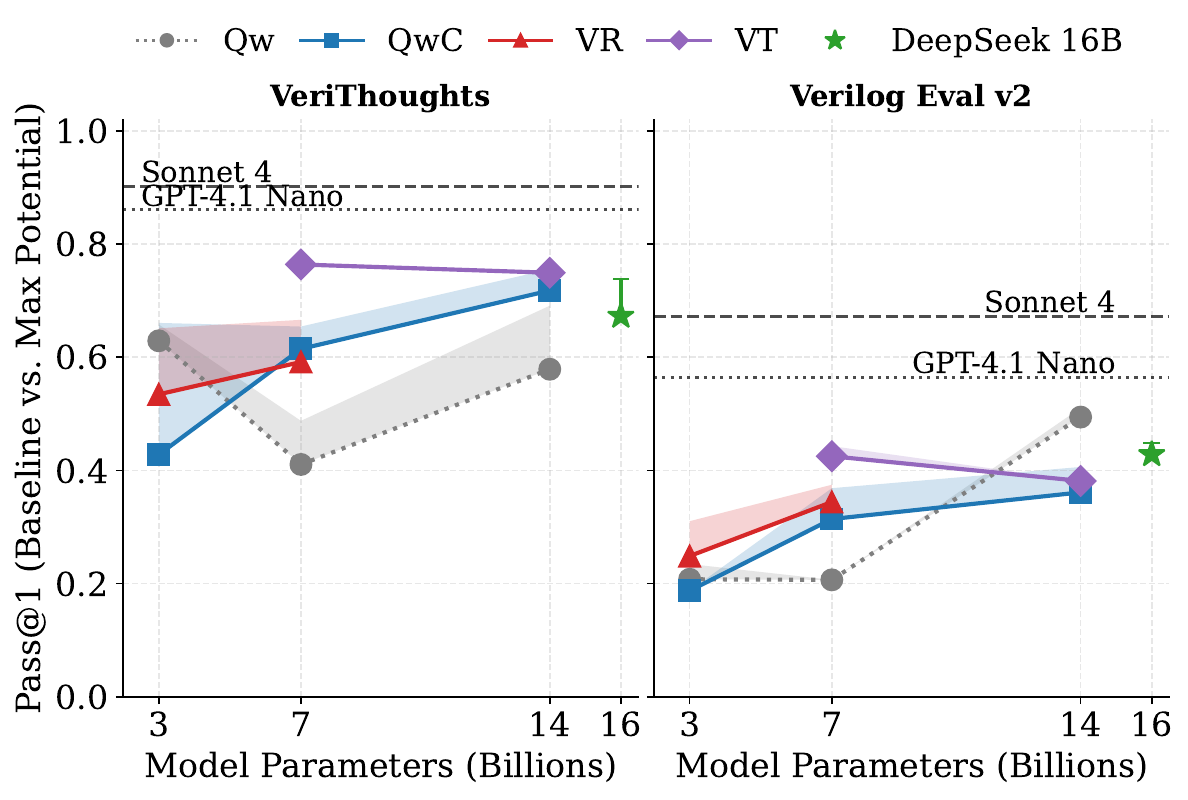}
    \caption{RQ1: Scaling and specialization for Verilog generation. 
Pass@1 normalized between each model’s baseline prompt performance and its best-performing prompting configuration, plotted against parameter count for two benchmarks: \textsc{VeriThoughts} (left) and \textsc{Verilog Eval v2} (right). 
Curves show the Qwen family (Qw), Qwen-Coder (QwC), and Verilog-adapted models (VR and VT); Star denotes DeepSeek-CoderV2-16B. Horizontal reference lines correspond to selected commercial models.}
    \label{fig:scaling}
\end{figure}

\Cref{fig:scaling} highlights that both scale and specialization can improve single-shot correctness, but with distinct behaviors for VR and VT models. We include Claude Sonnet 4 and GPT-4.1 nano as commercial reference points representing the highest- and lowest-cost regimes among our selected proprietary models. On \textsc{VeriThoughts} (left), the VT models remain near the top of the open-source curves but exhibit mild degradation from 7B to 14B, suggesting that supervised fine-tuning on limited data provides strong performance at moderate scale but may not fully exploit increased capacity. VeriReason (VR), in contrast, improves from 3B to 7B but offers only marginal gains over Qwen-Coder at comparable sizes. On \textsc{Verilog Eval v2} (right), VR benefits from scaling more consistently and is competitive with, or exceeds, general-purpose families at similar sizes—consistent with being optimized directly against simulation-based pass criteria. The divergent behaviors of VT and VR across benchmarks underscore the importance of evaluating specialization under multiple benchmarks: models might be overly fit to a specific benchmark design. We also observe that the base Qwen family exhibits a non-monotonic scaling trend on \textsc{VeriThoughts} (3B\,$>$\,7B), while on \textsc{Verilog Eval v2} the Qw-14 model is the strongest among open models, approaching GPT-4.1n.

\vspace{.8em}
\noindent\textbf{Key Takeaway for RTL Workflows:} For local deployments, model scale remains the primary driver of baseline Verilog generation. While prompting strategies offer limited improvements, fine-tuning is highly effective provided the model size matches the data volume at hand. Users should maximize their local model capacity and exhaust prompting options before carefully sizing a fine-tuning pipeline to their dataset.

\subsection{RQ2: Sensitivity to prompting strategies}

\begin{figure}[t]
    \centering
    \includegraphics[width=\linewidth]{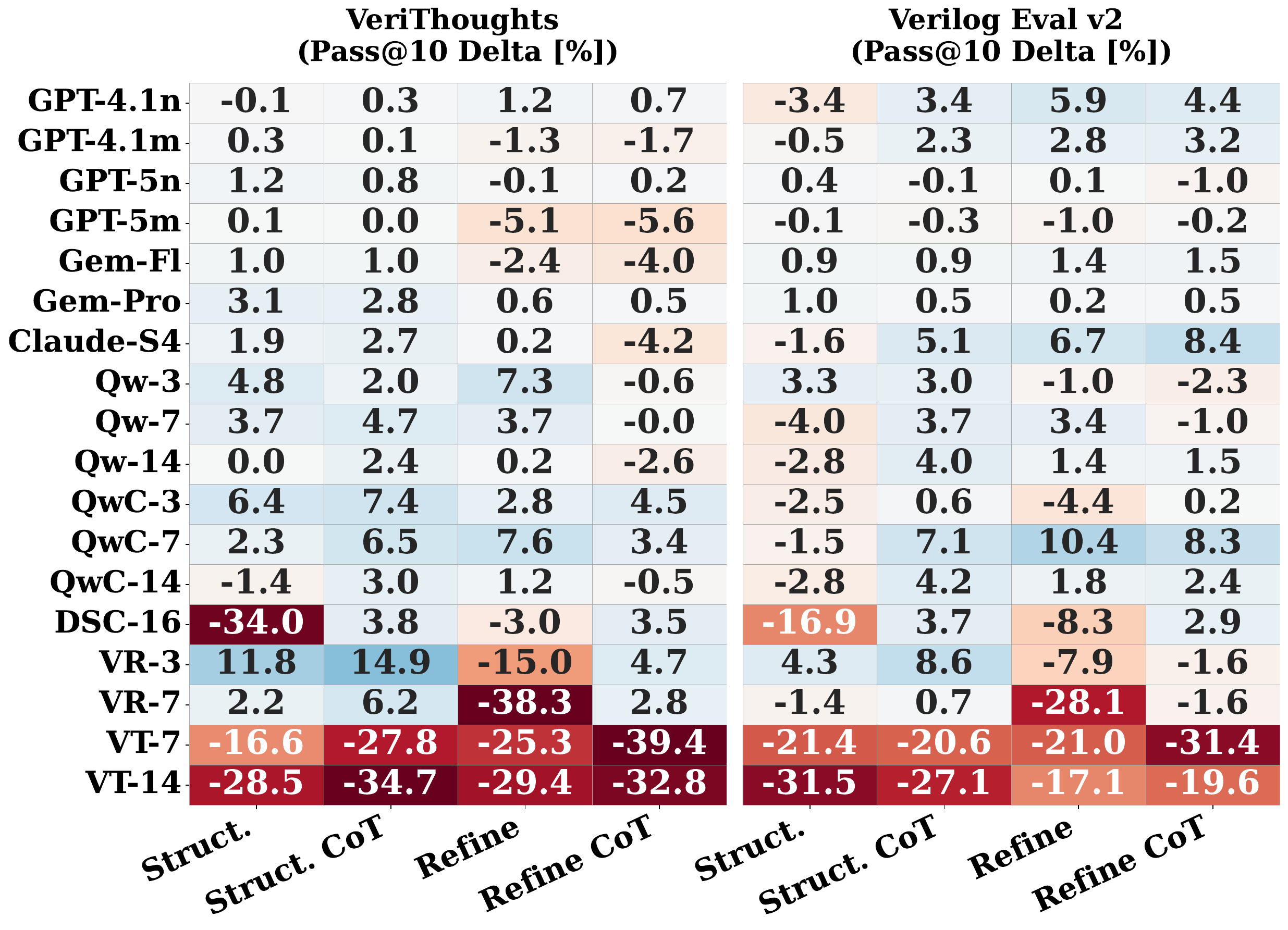}
    \caption{RQ2: Sensitivity to prompting strategies across LMs. Each cell reports the change in Pass@10 (percentage points) relative to the baseline prompt ({Base}), for four prompting variants ({Struct}, {Struct CoT}, {Refine}, {Refine CoT}) evaluated without ICL. Results are shown for \textsc{VeriThoughts} (left) and \textsc{Verilog Eval v2} (right); blue indicates gains and red indicates degradations.}
    \label{fig:heatmap}
\end{figure}

\Cref{fig:heatmap} shows that prompt sensitivity is highly dependent on model and benchmark. Structured prompting typically yields consistent gains for many open-source models (especially Qwen-Coder variants), while adding CoT is mixed: it helps some models but can degrade performance for others. Prompt refinement is the least stable intervention: several models exhibit large negative deltas under \textsc{Refine} and \textsc{Refine CoT}, indicating that rewriting the specification can introduce failure modes that outweigh any benefits.

A notable exception are the VT models, which perform well on straightforward Verilog generation (e.g., baseline), but degrade when the prompting requires extra capabilities such as structured output, specification rewriting, or explicit intermediate reasoning. This suggests that the VT supervised fine-tuning may have traded off some general instruction-following ability in favor of generation on its target distribution. In contrast, VR is comparatively capable of following auxiliary instructions (e.g., rewriting and explicit reasoning) and benefits from structured prompts.

\begin{figure*}[th]
    \centering
    \includegraphics[width=\linewidth]{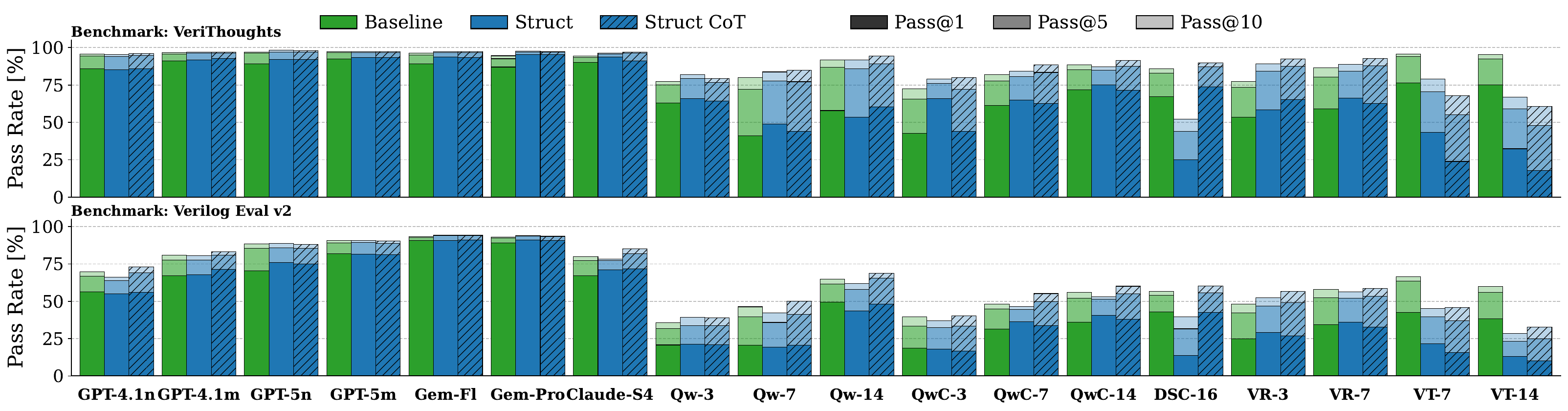}
    \caption{RQ2: Impact of structured output prompting. For each LM, we report P@1/P@5/P@10 under baseline (Baseline), structured-output variant (Struct), and chain-of-thought extension (Struct CoT). Results for \textsc{VeriThoughts} (top) and \textsc{VerilogEval v2} (bottom).}
    \label{fig:struct_impact}
\end{figure*}
\textbf{Effect of structured prompting.} \Cref{fig:struct_impact} shows that a structured output format is not uniformly beneficial: for the Qwen family, structured prompting alone often degrades performance, while adding a lightweight CoT stage on top of baseline tends to recover and even improve results. More broadly, the gains from structure are strongest for smaller, code-specialized models (e.g., QwC variants), while commercial models are less sensitive (smaller deltas).

A key exception is the VeriThoughts-tuned models (VT-7/VT-14), which exhibit substantial regression under structured prompting on both benchmarks, suggesting that they may be less robust to prompts that add non-generation constraints (e.g., explicit structure and intermediate reasoning) and are best used with minimal prompt transformations.

\begin{figure*}[thb]
    \centering
    \includegraphics[width=\linewidth]{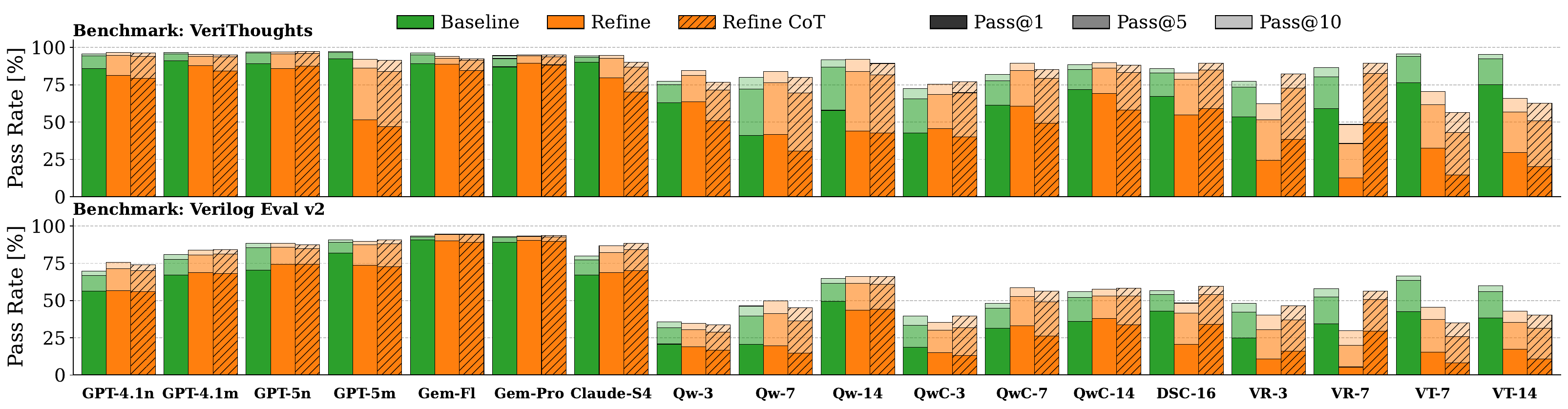}
    \caption{RQ2: Impact of prompt refinement. For each LM, we report P@1/P@5/P@10 under baseline (Baseline) vs two-stage refinement pipeline (Refine) and chain-of-thought variant (Refine CoT). Results for \textsc{VeriThoughts} (top) and \textsc{VerilogEval v2} (bottom).}
    \label{fig:refine_impact}
\end{figure*}
\textbf{Effect of prompt refinement.} \Cref{fig:refine_impact} shows that refinement is the most brittle inference-time intervention we evaluate. Across both benchmarks, many strong reasoning and Verilog-specialized models degrade  (especially in P@1 rate) when asked to rewrite the specification before producing RTL (e.g., GPT-5m and the VT family). For GPT-5m, we found that it tends to "over deliver" proposing a testbench that causes the evaluation to fail, dropping the P@1 score. We found that post-processing the output to remove the testbench makes it perform in line with the other experiments. For VT, the model appears to have reduced capabilities in following tasks outside of code generation. In contrast, some smaller open models  benefit from refinement (notably Qw-3 on \textsc{VeriThoughts} and QwC-7 on \textsc{Verilog Eval v2}), where rewriting resolves underspecified requirements and improve single-shot correctness.

Across both benchmarks, adding CoT to refinement (Refine CoT) is mixed: it occasionally recovers performance by making the rewrite explicit and self-consistent, but it also increases prompt length and introduces degrees of freedom that can amplify specification drift. Overall, these results indicate that prompt refinement is not a ``safe'' default for Verilog: it should be applied selectively (e.g., specifications are ambiguous) and validated for the target task.


\textbf{Effect of chain-of-thought.} 
The impact of adding an explicit CoT stage can be observed in \Cref{fig:struct_impact,fig:refine_impact}, which together show that CoT yields heterogeneous effects depending on both the model and the surrounding prompt template. Under structured output prompting (Struct), CoT often acts as a lightweight ``planning'' step that can partially mitigate format-induced errors and improve higher-$k$ pass rates, particularly on \textsc{Verilog Eval v2}. However, under \textsc{VeriThoughts}, the same added reasoning step can be neutral or even harmful, introducing extra, unverified assumptions that propagate into RTL. 

\begin{figure*}[thb]
    \centering
    \includegraphics[width=\linewidth]{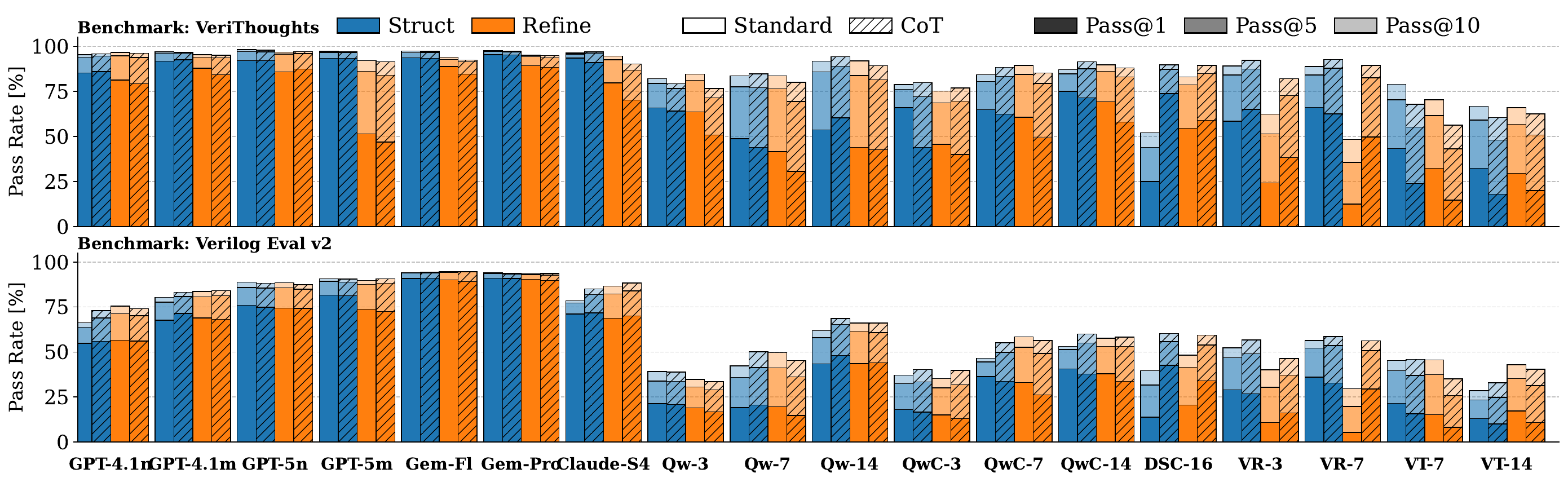}
    \caption{RQ2: Impact of chain-of-thought prompting. For each model, we compare two base prompts (structured output prompting, Struct; and prompt refinement, Refine) with and without an explicit intermediate reasoning stage (CoT). Results are shown for \textsc{VeriThoughts} (top) and \textsc{Verilog Eval v2} (bottom), reporting P@1/P@5/P@10.}
    \label{fig:cot_impact}
\end{figure*}

\Cref{fig:cot_impact}  isolates the effect of CoT. For refinement-based prompting (Refine), CoT is less consistently helpful because refinement already introduces an additional transformation step; adding CoT increases the risk of specification drift, which can outweigh any gains from additional deliberation. Overall, these results suggest that CoT should be treated as a prompt-conditional tool: it is most useful as a stabilizer for structured templates, but it is not a universally beneficial add-on for high-assurance specification-to-RTL generation.

\begin{figure*}[thb]
    \centering
    \includegraphics[width=\linewidth]{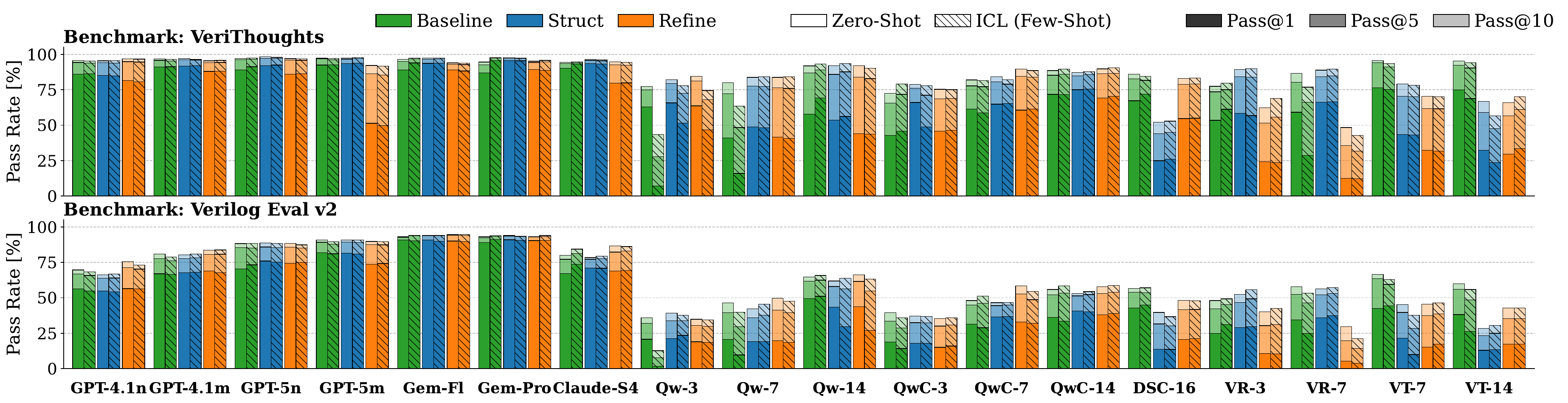}
    \caption{RQ2: Impact of ICL. For each model and prompt family (Baseline, Struct, and Refine), we compare zero-shot prompting against few-shot ICL, reporting P@1/P@5/P@10 on \textsc{VeriThoughts} (top) and \textsc{Verilog Eval v2} (bottom).}
    \label{fig:icl}
\end{figure*}
\textbf{Effects of ICL.} \Cref{fig:icl} shows that ICL is not a uniformly ``safe'' improvement for Verilog generation. When paired with a stable prompt template (especially structured prompting), few-shot exemplars often improve higher-$k$ success rates for smaller open-source models. That is, examples constrain output format and highlight common Verilog idioms. However, for baseline prompts for smaller models and several Verilog-specialized models, ICL can be neutral or even harmful, likely because longer contexts and exemplar-specific conventions increase sensitivity to prompt mismatch and semantic drift. Overall, ICL is best treated as a model- and template- dependent tool rather than a default setting, and should be validated for the target use case.

\begin{figure}[h]
    \centering
    \includegraphics[width=\linewidth]{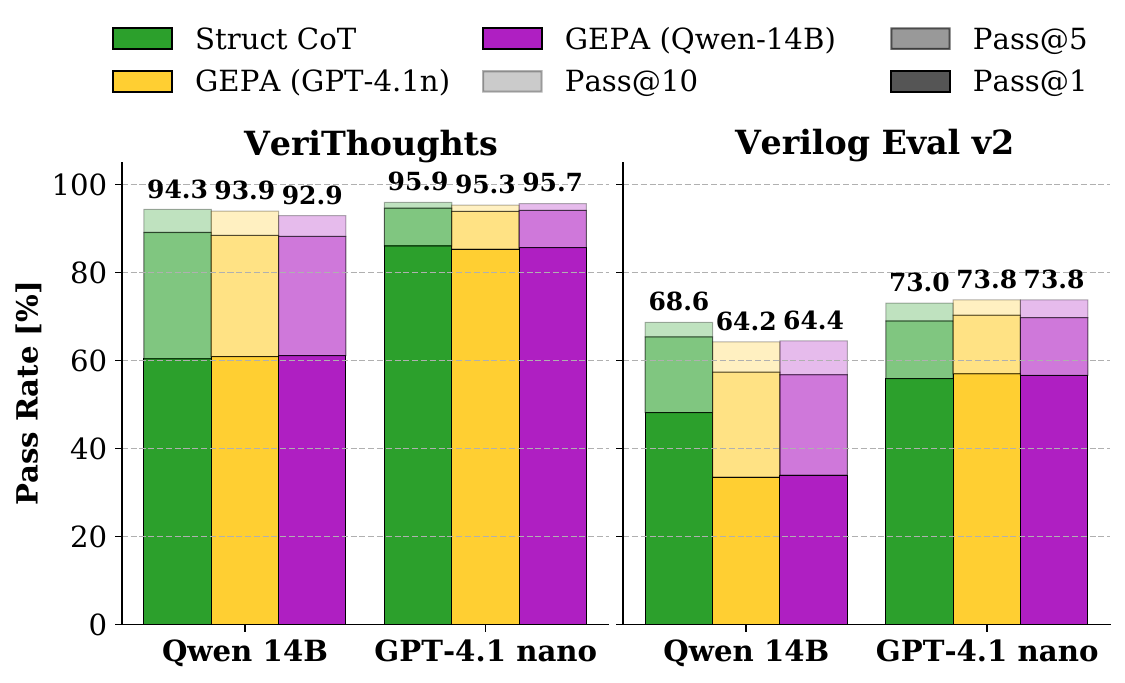}
    \caption{RQ2: Prompt optimization (GEPA). We compare a strong baseline prompt (Struct CoT) with GEPA-optimized prompts for two representative models (Qwen2.5-14B and GPT-4.1 nano), reporting P@1, P@5, and P@10 on \textsc{VeriThoughts} (left) and \textsc{Verilog Eval v2} (right).}
    \label{fig:gepa}
\end{figure}

\textbf{Effects of prompt optimization with GEPA.}
\Cref{fig:gepa} shows that GEPA performs within the margin of variability of our non-optimized prompting strategy (Struct CoT) for both representative LMs. On both benchmarks, the GEPA-optimized prompts achieve performance that is comparable to Struct CoT across Pass@1/5/10, with differences that are small and inconsistent in direction.

We observe qualitative differences in the learned prompts. As reported in~\Cref{app:gepa_prompts}, the prompt optimized for GPT-4.1 nano is short and appears incomplete; however, re-running GEPA with a different random seed produced a similarly short prompt and comparable performance, suggesting this outcome is not a one-off artifact of the optimizer initialization. In contrast, the optimized prompt for Qwen2.5-14B is substantially more verbose and appears ``complete'' but still does not translate into improved pass rates.

Overall, we found GEPA relatively lightweight to run and believe it remains a sensible method to try, especially given the strong results reported in the original GEPA paper~\cite{agrawal2025gepareflectivepromptevolution}. However, our experiments show clear limitations, and prompt evolution is not a universal win, highlighting the challenges in specification-to-RTL generation. 

\vspace{0.8em}
\noindent\textbf{Key Takeaway for RTL Workflows:} Inference-time prompt engineering is a cost-effective ``first step'' that can significantly narrow the gap between off-the-shelf LLMs and custom hardware models. However, it operates under a strict capacity ceiling. For the complex spatial and temporal concurrency constraints inherent in HDLs, no amount of prompt optimization can artificially confer hardware design capabilities the base model lacks; domain-specific fine-tuning remains indispensable for production-grade RTL generation.

\subsection{RQ3: Verilog fine-tuning vs. strong prompting}
Now that we have presented all results, we can compare what is achievable with inference-time adaptations alone to what is achievable with training-time adaptations.
\begin{figure}[h]
    \centering
    \includegraphics[width=\linewidth]{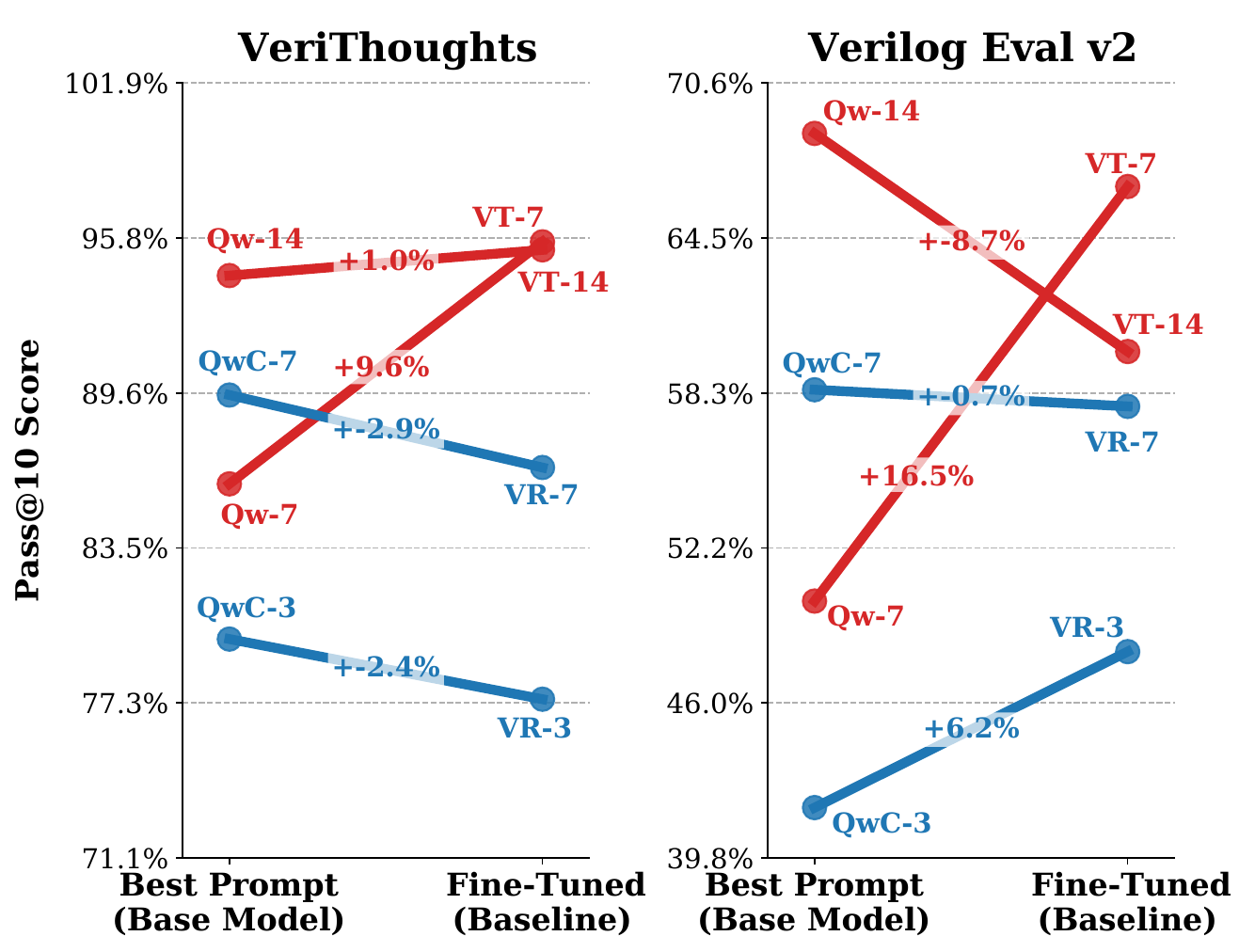}
    \caption{RQ3: We compare fine-tuned models with the baseline prompt against their base models with their strongest prompting technique, reporting P@10 on \textsc{VeriThoughts} (left) and \textsc{Verilog Eval v2} (right). }
    \label{fig:ft_vs_prompt}
\end{figure}
\Cref{fig:ft_vs_prompt} highlights that ``strong prompting'' can narrow the gap to fine-tuning for some base models, but the relative benefit depends on both the model family and the benchmark. On \textsc{VeriThoughts}, prompt engineering yields large gains for some Qwen variants (e.g., Qw-7 and Qw-14), while fine-tuning provides a more consistent improvement for Verilog-specialized models (VT-7/VT-14). In contrast, the VeriReason models (VR-3/VR-7) show limited uplift from fine-tuning under \textsc{VeriThoughts} compared to their best-prompted baseline models. This suggests that fine-tuning for a benchmark does not ensure generalizable gains.

On \textsc{Verilog Eval v2}, the picture shifts: fine-tuning (especially for VR-3 and VT-7) produces substantial improvements relative to the best prompted base models, consistent with these models being adapted (directly or indirectly) to the benchmark. Overall, the figure suggests that inference-time prompt optimization is a strong and inexpensive first step, but it is not a complete substitute for training-time adaptation; when the target distribution matches the fine-tuning objective, specialized models can achieve gains that prompting alone does not reliably recover.

\vspace{0.8em}
\noindent\textbf{Key Takeaway for RTL Workflows:} 
The optimal prompting strategies identified for a base model often become detrimental once that model undergoes fine-tuning. While structured prompting and intermediate reasoning successfully steer general-purpose models toward valid Verilog, they actively harm models fine-tuned specifically for hardware. Because fine-tuning rigidly narrows a model's output distribution for direct code generation, forcing specialized models to produce intermediate results or adhere to out-of-distribution structures can be detrimental. 

\subsection{RQ4: Stability across benchmarks}

\begin{figure}[h]
    \centering
    \includegraphics[width=\linewidth]{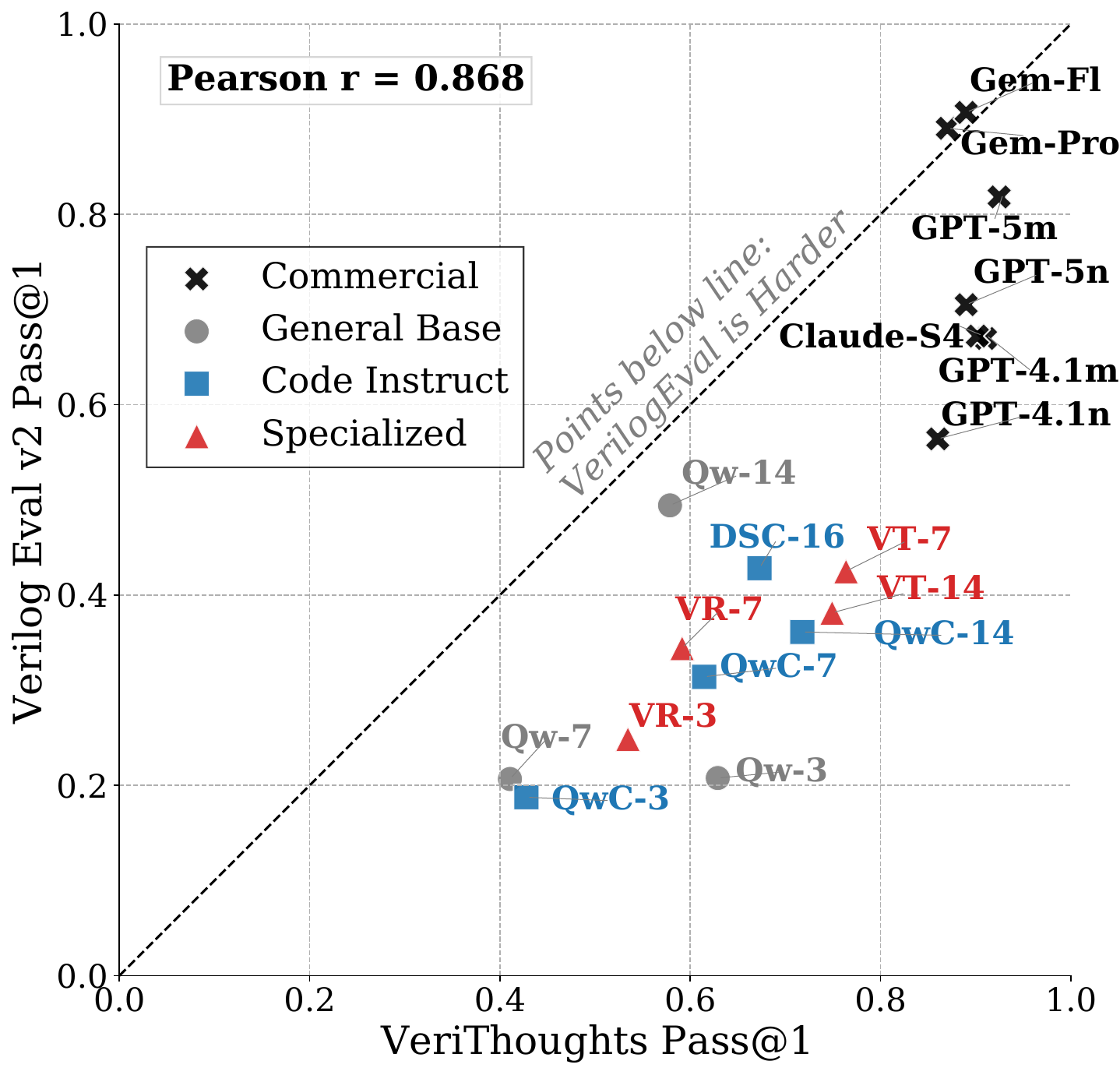}
    \caption{Correlation between the two benchmarks. Each point is an LM, plotting Pass@1 on \textsc{VeriThoughts} (x-axis) versus \textsc{Verilog Eval v2} (y-axis). Deviations from the diagonal reveal benchmark-dependent strengths.}
    \label{fig:benchmark_correlation}
\end{figure}

\Cref{fig:benchmark_correlation} shows a strong positive correlation between the two benchmarks ($r=0.868$), suggesting that LM ranking is broadly stable across benchmarks. However, most points lie below the $y=x$ line, indicating that \textsc{Verilog Eval v2} is systematically harder (lower Pass@1) than \textsc{VeriThoughts} for a large fraction of LMs. This gap is consistent with testbench-driven evaluation, which penalizes a different set of failure modes (e.g., mismatched I/O timing, reset conventions, or behavioral edge cases) than functional equivalence against a golden RTL and the benchmarks' design.

The outliers are particularly informative. Once again, we have evidence that the Verithought models are over-tuned to their own benchmarks. We find two general LMs, Qw-14 and DSC-16, on the Pareto for the SLMs together with VT-7. This highlights how SLMs can remain effective without additional fine-tuning. 
Conversely, the commercial models cluster near the top-right, remain strong across both benchmarks, and exhibit smaller benchmark gaps. Gemini models are the only two to score higher on \textsc{Verilog Eval v2} than \textsc{VeriThoughts}. 
Overall, these results argue against relying on a single benchmark: while the two benchmarks agree on coarse trends, the residual disagreement is large enough to change conclusions about which models and adaptation strategies are reliable for a given deployment setting.

\vspace{0.8em}
\noindent\textbf{Key Takeaway for RTL Workflows:} Evaluating a model on a single benchmark is viable only when the intended deployment strictly aligns with that specific task. For broader hardware design applications and use cases, diverse multi-benchmark evaluation is crucial. This comprehensive testing ensures that while a model improves on a targeted capability, its performance does not simultaneously degrade into unusability on other essential tasks.

\section{Discussion}

\subsection{Verilog as a distinct code-generation regime.}
Our results suggest that Verilog generation is not merely a variant of conventional code generation, but a qualitatively different task regime for language models. In software-centric settings, partial correctness and iterative refinement are often acceptable, and many inference-time techniques reliably improve performance. In contrast, RTL generation imposes tighter global constraints on semantics, structure, and timing, where small deviations can invalidate an otherwise plausible solution. This difference may help explain why some prompting strategies that work well in software settings are less reliable in hardware tasks.

For instance, our evolutionary prompt optimization (GEPA) experiments yielded marginal and inconsistent improvements across the evaluated models. To investigate whether this was an artifact of the search parameters, we scaled up the optimization effort, increasing the number of samples and iterations. While this produced qualitatively more detailed and ``better-looking'' prompts tailored to the structural nature of Verilog, it did not translate to quantitative improvements in benchmark scores. We theorize that this exposes a fundamental capacity ceiling: if a model lacks the underlying latent representations or parameter scale to comprehend complex EDA concepts, prompt engineering can only take it so far. Analogous to explaining a quantum mechanics problem to a novice, where no amount of instructional clarity can substitute for foundational knowledge, optimizing \textbf{prompts cannot artificially induce hardware design capabilities that the base model simply does not possess}. For the EDA community, this implies that \textbf{future productivity gains will require advancing the pre-training and domain-specific fine-tuning paradigms of the models} themselves, rather than relying solely on inference-time prompt wrappers.

\subsection{Model--prompt interactions drive performance.}
Across RQ1–RQ4, a consistent theme is that performance is governed by interactions between model properties and inference-time scaffolding rather than by any single factor. Scaling and specialization can both improve performance (RQ1), but specialization is only beneficial when aligned with the evaluation and task distribution, and the model size matches the available data volume. Crucially, the optimal prompting strategy shifts based on a model's optimization state: while structured reasoning and inference-time scaffolding successfully steer general-purpose LMs, these same techniques actively harm models fine-tuned specifically for hardware, forcing them out-of-distribution (RQ2 and RQ3). Finally, evaluating on a single benchmark is only viable for narrowly targeted deployments; broader hardware use cases require diverse, multi-benchmark validation to ensure that optimizing for one task does not cause catastrophic degradation in others (RQ4).

\subsection{Specialization--robustness trade-offs.}

The behavior of Verilog-tuned reasoning models highlights both the benefits and trade-offs of specialization. VT-7 achieves strong performance on its specialized task but degrades under modest prompt transformations. This suggests that VT-7 has learned a narrower, highly optimized mapping for its training distribution, at the cost of reduced robustness to distributional shift. Specifically, we attribute this to the fact that the VT models are fine-tuned using Reinforcement Learning (RL) examples that rigidly enforce a specific input-output format. By learning to reply only in that exact format, the model overfits to the structure and experiences catastrophic forgetting of its broader instruction-following capabilities. 

In contrast, VR-7 exhibits lower peak performance but more stable behavior across prompting variations. Such specialization--robustness trade-offs are consequential for deployment, where prompt formats, toolchains, and workflows inevitably vary. Future fine-tuning methodologies for hardware LLMs must focus on improving Verilog generation accuracy without constraining the model's underlying flexibility, potentially through data-mixing strategies that preserve general instruction-following capabilities while enabling domain-specific HDL tasks.

\subsection{Scope and limitations.}
Our goal is to provide an empirical map of how model class and prompt structure interact for Verilog generation, under a single, controlled evaluation protocol. Accordingly, we fix decoding to a single temperature to isolate prompt effects and enable consistent cross-model comparisons; exploring broader decoding regimes and per-model tuning is a natural extension. Finally, our analysis emphasizes aggregate reliability metrics (pass@k) across benchmarks rather than a full taxonomy of error modes; targeted failure analyses (e.g., reset/clocking semantics, interface mismatches, and near-miss equivalence/simulation failures) would complement the trends we report.

\subsection{Takeaways}

Our empirical evaluation of 18 language models across diverse prompting strategies yields a practical decision framework for Electronic Design Automation (EDA) teams seeking to integrate LLMs into Verilog workflows. Practitioners should navigate the space of scale, specialization, and inference-time compute as follows:

\begin{itemize}
    \item \textbf{Assess IP constraints and model scale:} If proprietary Intellectual Property (IP) restrictions permit sending RTL data to cloud APIs, teams should default to large-scale commercial models (e.g., GPT-4 class) and apply advanced prompt engineering (such as Chain-of-Thought and iterative refinement), which generally scales well with model size.
    \item \textbf{Maximize local capacity if cloud is restricted:} If strict IP confidentiality requires local, on-premise deployment, engineers should deploy the largest open-weight model their hardware constraints allow (e.g., 7B to 70B parameters) and exhaust inference-time prompt interventions.
    \item \textbf{Cross the capacity ceiling via fine-tuning:} If prompt-engineered base models fail to deliver the required syntactical and semantic correctness, practitioners have hit the ``capacity ceiling'' of inference-time optimization. At this stage, teams must pivot from prompt engineering to domain-specific fine-tuning (e.g., adopting or training models like VR-7 or VT-7) to embed strict Verilog structural constraints directly into the model weights. 
    \item \textbf{Anchor deployment in robust benchmarking:} Regardless of the chosen path, reliance on a single evaluation suite is a critical vulnerability. Deployment must be continuously validated against benchmarks that test both simulation-based correctness and formal equivalence to prevent over-optimization on narrow task distributions.
\end{itemize}

\subsection{Future Work}
The limitations uncovered in this study highlight several immediate directions for future research in ML-driven hardware design:

\textbf{1. Preserving Robustness via Data-Mixing:} Our analysis of highly specialized Verilog SLMs (such as the VeriThoughts family) revealed a vulnerability to structural prompting, likely caused by rigid input-output formatting during fine-tuning. Future hardware fine-tuning methodologies must investigate data-mixing strategies---combining strict HDL generation tasks with diverse, general-purpose instruction-following data---to improve Verilog accuracy without triggering catastrophic forgetting of the model's baseline flexibility.

\textbf{2. Hardware-Native Optimization Pipelines:} General-purpose prompt optimization techniques (such as our GEPA trials) yielded marginal improvements, exposing the limits of applying software-centric algorithms to hardware description languages. Just as one cannot successfully prompt a novice to solve a quantum mechanics problem without foundational knowledge, optimization algorithms cannot induce complex spatial and temporal concurrency constraints if the base model lacks the latent representations. Future work should focus on hardware-native optimization that structurally aligns with EDA intermediate representations (IR) rather than plain text. 

\textbf{3. System-Level Verification Benchmarks:} As models saturate existing module-level benchmarks like VerilogEval, the community requires more rigorous evaluation frameworks. Future benchmarks must move beyond isolated, single-module generation and test system-level integration, bus protocol compliance (e.g., AXI/APB), and multi-module timing constraints to better reflect the realities of modern silicon tape-outs.




\section{Conclusion}
We presented a comprehensive empirical study of LLM-driven Verilog generation, evaluating 18 models across diverse inference-time prompting strategies and two rigorous hardware benchmarks. Our results demonstrate that Verilog generation is a distinct, highly constrained task regime where conventional software-centric prompt engineering often hits a fundamental ``capacity ceiling.'' While advanced prompting can narrow the performance gap for general-purpose models, it cannot artificially induce hardware design capabilities if the base model lacks the requisite latent representations. 

Furthermore, we highlighted a critical specialization--robustness trade-off: while domain-specific fine-tuning is essential for capturing the strict spatial and temporal concurrency of hardware, current specialized Small Language Models (SLMs) often overfit to rigid training formats, resulting in fragility under complex prompting. For EDA practitioners, our findings provide a practical deployment framework: leverage large commercial models with advanced prompting when IP constraints allow, but invest in robust, data-mixed fine-tuning for secure, local deployment. Ultimately, this work underscores that advancing ML for hardware design requires moving beyond inference-time text wrappers toward hardware-native optimization, better-aligned training signals, and system-level verification benchmarks.

\appendix

\section*{System Prompts}
\label{app:sys_prompts}

This section lists the system-level prompts used to define the behavior of the core agents in our pipelines.
\begin{promptbox}[title=\textbf{VerilogWriter} system prompt]
You are an expert Verilog RTL engineer. 
Your task is to generate a complete, correct, 
and synthesizable Verilog module implementation from user specification. 
The top-level module should be named according to the specified name.
\end{promptbox}

\begin{promptbox}[title=\textbf{SpecRefiner} system prompt]
You are an expert Verilog hardware engineer. Your task is to refine and clarify a given Verilog module specification to ensure it is complete, unambiguous, and suitable for implementation. The specification will be given to an Intern, who will implement the Verilog module based on your refined version. The refined specification should address any potential ambiguities, fill in missing details, and enhance clarity while preserving the original intent.
\end{promptbox}

\section*{GEPA-Optimized System Prompts}
\label{app:gepa_prompts}

This section lists the system prompts obtained through GEPA prompt optimization for selected models.

\begin{promptbox}[title=\textbf{VerilogWriter} system prompt (optimized with GEPA for GPT-4.1n)]
\begin{Verbatim}[fontsize=\footnotesize, breaklines, breakanywhere]
verilog ... ```.

```
Read the `specification` carefully to identify the module name, interface requirements, and logic function.
Produce the `reasoning` and the `verilog_code`.
\end{Verbatim}
\end{promptbox}
\vspace{-0.3em}

\begin{figure*}
\begin{promptbox}[title=\textbf{VerilogWriter} system prompt (optimized with GEPA for Qw-14)]
\begin{Verbatim}[fontsize=\footnotesize, breaklines, breakanywhere]
You are an expert Verilog HDL designer and engineer. Your task is to interpret natural language specifications for digital logic circuits and produce the corresponding Verilog code.
**Input Format:**
You will be provided with a field named `specification` containing the design requirements, interface definitions (inputs/outputs), and functional behavior.
**Output Format:**
You must produce two fields:
1. `reasoning`: A detailed breakdown of the design logic, identifying inputs/outputs, determining whether the logic is combinational or sequential, and planning the Verilog syntax strategies (e.g., choice of operators, control structures).
2. `verilog_code`: The valid, synthesizable Verilog code within a markdown code block (```verilog ... ```).
**Crucial Guidelines and Domain-Specific Constraints:**
1.  **Synchronous vs. Asynchronous Resets:**
    *   Pay extreme attention to the type of reset requested.
    *   **Synchronous Reset:** Do **not** include `posedge reset` in the sensitivity list. The sensitivity list should only contain the clock (e.g., `always @(posedge clk)`). Check the reset condition *inside* the block.
    *   **Asynchronous Reset:** Include the reset in the sensitivity list (e.g., `always @(posedge clk or posedge reset)`).
2.  **Signal Declarations (`reg` vs `wire`):**
    *   If a signal is assigned values inside an `always` block (procedural assignment), it **must** be declared as `reg` (e.g., `output reg [3:0] out`).
    *   If a signal is assigned using the `assign` keyword (continuous assignment), it is treated as a wire and does not need the `reg` keyword.
3.  **Combinational vs. Sequential Logic:**
    *   If the specification asks for a "combinational circuit" (e.g., a simple multiplier or logic gate block), do not use clocked `always` blocks or registers, even if a clock signal is present in the interface list. Use `assign` statements or `always @(*)` logic.
    *   For combinational logic inside an `always @(*)` block, use blocking assignments (`=`).
    *   For sequential logic inside a clocked `always` block, use non-blocking assignments (`<=`).
4.  **Interface Accuracy:**
    *   Adhere strictly to the naming conventions and bit-widths provided in the specification.
    *   Do not hallucinate inputs (like `clk` or `reset`) if they are not specified, unless the persona instructions explicitly imply standard interfaces for specific module types. Conversely, if inputs are provided in the list, ensure they are included in the module definition even if unused (though usually, they should be used).
\
5.  **Syntax:**
    *   Ensure all modules are closed with `endmodule`.
    *   Avoid syntax errors related to port declarations (e.g., declaring a port as both input and reg in strict ANSI style vs non-ANSI style). Preferred style is ANSI: `module name (input clk, output reg out);`.
\end{Verbatim}
\end{promptbox}
\vspace{-1.2em}
\end{figure*}

\section*{Numerical Results}
\label{app:num_results}

This section reports numerical results for all models and prompting conditions. Results are split into two tables: commercial models (\Cref{tab:results_comm}) and open-source models (\Cref{tab:results_oss}). The tables report pass@$k$ (with $k\in\{1,5,10\}$) on both benchmarks (\textsc{VeriThoughts} and \textsc{Verilog Eval v2}), with separate columns for runs \emph{without} in-context learning (No ICL) and \emph{with} in-context learning (ICL). Bolded values indicate the best result \emph{within each model} across prompting conditions, and underlined values indicate the best result \emph{within each model family group} (commercial vs. open-source).


\setlength{\tabcolsep}{2pt}
\setlength{\aboverulesep}{0.4em}
\setlength{\belowrulesep}{0.6em}
    
\begin{table*}[h] 
\centering
\caption{Results for commercial models across prompting conditions.}
\label{tab:results_comm}

\begin{tabular}{
  ll
  @{\hspace{0.4em}\vrule width 1.2pt \hspace{0.4em}} ccc
  @{\hspace{0.4em}\vrule width 0.2pt \hspace{0.4em}} ccc
  @{\hspace{0.4em}\vrule width 1.2pt\hspace{0.4em}} ccc
  @{\hspace{0.4em}\vrule\hspace{0.4em}} ccc
}

\specialrule{1.2pt}{0em}{0.4em}

\multirow{3}{*}{\textsc{Model}} & \multirow{3}{*}{\textsc{Exp.}} & \multicolumn{6}{c}{\textsc{VeriThoughts}} & \multicolumn{6}{c}{\textsc{Verilog Eval v2}} \\

 &  & \multicolumn{3}{c}{\textsc{No ICL}} & \multicolumn{3}{c}{\textsc{ICL}} & \multicolumn{3}{c}{\textsc{No ICL}} & \multicolumn{3}{c}{\textsc{ICL}} \\

 &  & \textsc{P@1} & \textsc{P@5} & \textsc{P@10} & \textsc{P@1} & \textsc{P@5} & \textsc{P@10} & \textsc{P@1} & \textsc{P@5} & \textsc{P@10} & \textsc{P@1} & \textsc{P@5} & \textsc{P@10} \\

\specialrule{1.2pt}{0.2em}{0.2em}

\multirow{5}{*}{GPT-4.1n} & Base & 86.0\% & 94.2\% & 95.6\% & 86.3\% & 93.8\% & 95.3\% & 56.4\% & 66.8\% & 69.6\% & 54.7\% & 65.7\% & 68.2\% \\
 & Struct & 85.2\% & 94.1\% & 95.5\% & 84.8\% & 93.7\% & 95.5\% & 54.9\% & 63.7\% & 66.3\% & 54.3\% & 64.0\% & 66.9\% \\
 & Struct CoT & 86.0\% & 94.6\% & 95.9\% & \textbf{86.4\%} & \textbf{95.3\%} & \textbf{96.8\%} & 55.9\% & 69.0\% & 73.0\% & 56.0\% & 69.4\% & 73.9\% \\
 & Refine & 81.4\% & 94.9\% & 96.7\% & 80.5\% & 94.5\% & 96.7\% & \textbf{56.5\%} & \textbf{71.3\%} & \textbf{75.5\%} & 56.3\% & 70.3\% & 73.0\% \\
 & Refine CoT & 79.3\% & 93.9\% & 96.2\% & 78.4\% & 93.0\% & 95.4\% & 56.1\% & 70.1\% & 74.1\% & 56.2\% & 70.4\% & 74.6\% \\
 
\specialrule{1.2pt}{0.2em}{0.2em}

\multirow{5}{*}{GPT-4.1m} & Base & 91.0\% & 95.7\% & 96.8\% & 91.3\% & 95.4\% & 96.5\% & 67.0\% & 77.6\% & 80.9\% & 66.5\% & 76.0\% & 78.9\% \\
 & Struct & 91.8\% & 96.3\% & \textbf{97.0\%} & 92.0\% & 95.8\% & 96.4\% & 67.7\% & 77.6\% & 80.4\% & 67.9\% & 78.2\% & 81.0\% \\
 & Struct CoT & \textbf{92.7\%} & \textbf{96.4\%} & 96.9\% & \textbf{92.7\%} & 96.2\% & 96.9\% & \textbf{71.4\%} & 80.9\% & 83.2\% & 71.0\% & 80.0\% & 82.4\% \\
 & Refine & 88.0\% & 94.0\% & 95.5\% & 88.2\% & 94.3\% & 95.7\% & 68.9\% & 80.7\% & 83.7\% & 67.6\% & 80.8\% & 83.8\% \\
 & Refine CoT & 84.4\% & 93.6\% & 95.1\% & 84.4\% & 93.6\% & 95.2\% & 68.2\% & 81.3\% & 84.0\% & 68.9\% & \textbf{82.3\%} & \textbf{85.2\%} \\

\specialrule{1.2pt}{0.2em}{0.2em}

\multirow{5}{*}{GPT-5n} & Base & 89.0\% & 96.2\% & 97.0\% & 91.3\% & 96.5\% & 97.3\% & 70.5\% & 85.5\% & 88.3\% & 73.3\% & 85.1\% & 88.4\% \\
 & Struct & 92.1\% & \textbf{97.1\%} & \underline{\textbf{98.2\%}} & \textbf{92.4\%} & \textbf{97.1\%} & 97.9\% & \textbf{76.0\%} & \textbf{85.9\%} & \textbf{88.7\%} & 75.1\% & 85.3\% & 88.4\% \\
 & Struct CoT & 92.2\% & 96.9\% & 97.9\% & 91.9\% & 96.5\% & 97.1\% & 74.8\% & 85.5\% & 88.2\% & 75.2\% & 84.8\% & 87.2\% \\
 & Refine & 85.9\% & 95.8\% & 96.9\% & 86.2\% & 95.7\% & 96.8\% & 74.3\% & 85.7\% & 88.4\% & 74.9\% & 85.1\% & 87.4\% \\
 & Refine CoT & 87.5\% & 96.1\% & 97.2\% & 86.7\% & 95.8\% & 96.9\% & 74.2\% & 84.9\% & 87.3\% & 74.5\% & 85.2\% & 87.9\% \\
 
\specialrule{1.2pt}{0.2em}{0.2em}

\multirow{5}{*}{GPT-5m} & Base & 92.5\% & 96.8\% & 97.2\% & 92.6\% & 96.2\% & 97.1\% & \textbf{81.8\%} & \textbf{89.2\%} & \textbf{90.8\%} & 81.1\% & 87.9\% & 89.7\% \\
 & Struct & 93.4\% & 96.7\% & 97.3\% & \textbf{93.8\%} & \textbf{97.0\%} & \textbf{97.7\%} & 81.6\% & \textbf{89.2\%} & \textbf{90.8\%} & 80.8\% & 89.0\% & 90.7\% \\
 & Struct CoT & 93.3\% & 96.6\% & 97.2\% & 93.5\% & 96.8\% & 97.5\% & 81.3\% & 88.8\% & 90.5\% & 81.4\% & 87.8\% & 89.2\% \\
 & Refine & 51.4\% & 86.3\% & 92.2\% & 49.7\% & 85.3\% & 91.7\% & 73.7\% & 87.5\% & 89.8\% & 74.2\% & 87.3\% & 89.5\% \\
 & Refine CoT & 46.9\% & 84.0\% & 91.6\% & 47.7\% & 84.0\% & 91.2\% & 72.6\% & 88.2\% & 90.6\% & 72.0\% & 88.0\% & 90.6\% \\

\specialrule{1.2pt}{0.2em}{0.2em}

\multirow{5}{*}{Gem-Fl} & Base & 89.0\% & 94.9\% & 96.4\% & \textbf{93.9\%} & 96.7\% & \textbf{97.5\%} & 90.7\% & 92.8\% & 93.3\% & 90.3\% & 93.7\% & 94.1\% \\
 & Struct & 93.7\% & 96.6\% & 97.4\% & 93.8\% & \textbf{96.8\%} & \textbf{97.5\%} & 90.8\% & 93.9\% & 94.2\% & 89.8\% & 93.8\% & 94.1\% \\
 & Struct CoT & 93.5\% & 96.6\% & 97.4\% & 93.2\% & 96.5\% & 97.0\% & \textbf{90.9\%} & 93.9\% & 94.2\% & 90.4\% & 94.3\% & 94.7\% \\
 & Refine & 88.9\% & 92.8\% & 94.0\% & 88.3\% & 92.8\% & 93.7\% & 90.1\% & 94.3\% & 94.7\% & 89.5\% & 94.1\% & 94.7\% \\
 & Refine CoT & 84.7\% & 91.5\% & 92.4\% & 87.0\% & 91.1\% & 91.8\% & 89.2\% & 94.4\% & 94.7\% & 89.6\% & \underline{\textbf{94.6\%}} & \underline{\textbf{95.2\%}} \\

\specialrule{1.2pt}{0.2em}{0.2em}

\multirow{5}{*}{Gem-Pro} & Base & 87.0\% & 92.6\% & 94.5\% & \underline{\textbf{95.6\%}} & \underline{\textbf{97.5\%}} & \textbf{97.6\%} & 89.0\% & 92.3\% & 93.1\% & \underline{\textbf{91.0\%}} & 93.4\% & 93.8\% \\
 & Struct & 95.5\% & 97.1\% & \textbf{97.6\%} & 95.5\% & 97.0\% & 97.5\% & 90.9\% & \textbf{93.6\%} & \textbf{94.1\%} & 90.5\% & 93.2\% & 93.5\% \\
 & Struct CoT & 95.3\% & 96.9\% & 97.4\% & 95.1\% & 96.6\% & 97.1\% & 90.8\% & 93.3\% & 93.6\% & 90.2\% & 93.4\% & 93.8\% \\
 & Refine & 89.4\% & 94.3\% & 95.1\% & 88.8\% & 94.6\% & 95.8\% & 90.4\% & 92.9\% & 93.3\% & 90.5\% & 93.3\% & 94.0\% \\
 & Refine CoT & 88.3\% & 93.7\% & 95.0\% & 88.0\% & 94.1\% & 95.1\% & 89.8\% & 92.7\% & 93.6\% & 90.0\% & 92.9\% & 93.7\% \\
 
\specialrule{1.2pt}{0.2em}{0.2em}

\multirow{5}{*}{Claude-S4} & Base & 90.2\% & 93.5\% & 94.5\% & 93.1\% & 94.6\% & 95.1\% & 67.2\% & 77.2\% & 79.9\% & \textbf{73.6\%} & 81.2\% & 84.3\% \\
 & Struct & \textbf{93.6\%} & 95.7\% & 96.4\% & 93.3\% & 95.4\% & 96.2\% & 71.1\% & 77.1\% & 78.4\% & 70.9\% & 77.5\% & 79.4\% \\
 & Struct CoT & 91.1\% & 96.4\% & 97.1\% & 91.3\% & \textbf{96.6\%} & \textbf{97.5\%} & 71.7\% & 81.9\% & 85.0\% & 72.0\% & 82.8\% & 86.2\% \\
 & Refine & 79.8\% & 92.6\% & 94.7\% & 79.9\% & 92.2\% & 94.3\% & 68.8\% & 82.3\% & 86.6\% & 69.3\% & 83.1\% & 86.2\% \\
 & Refine CoT & 70.3\% & 86.8\% & 90.3\% & 69.3\% & 86.2\% & 90.0\% & 70.1\% & \textbf{84.0\%} & \textbf{88.3\%} & 67.5\% & 82.9\% & 87.9\% \\

\specialrule{1.2pt}{0.2em}{0.2em}

\end{tabular}
\end{table*}


\setlength{\tabcolsep}{3pt}
\setlength{\aboverulesep}{0.3em}
\setlength{\belowrulesep}{0.2em}

\begin{table*}[h] 
\centering
\caption{Results for open-source models across prompting conditions.}
\label{tab:results_oss}

\begin{tabular}{
  ll
  @{\hspace{0.4em}\vrule width 1.2pt \hspace{0.4em}} ccc
  @{\hspace{0.4em}\vrule width 0.2pt \hspace{0.4em}} ccc
  @{\hspace{0.4em}\vrule width 1.2pt\hspace{0.4em}} ccc
  @{\hspace{0.4em}\vrule\hspace{0.4em}} ccc
}

\specialrule{1.2pt}{0em}{0.2em}

\multirow{3}{*}{\textsc{Model}} & \multirow{3}{*}{\textsc{Exp.}} & \multicolumn{6}{c}{\textsc{VeriThoughts}} & \multicolumn{6}{c}{\textsc{Verilog Eval v2}} \\

 &  & \multicolumn{3}{c}{\textsc{No ICL}} & \multicolumn{3}{c}{\textsc{ICL}} & \multicolumn{3}{c}{\textsc{No ICL}} & \multicolumn{3}{c}{\textsc{ICL}} \\

 &  & \textsc{P@1} & \textsc{P@5} & \textsc{P@10} & \textsc{P@1} & \textsc{P@5} & \textsc{P@10} & \textsc{P@1} & \textsc{P@5} & \textsc{P@10} & \textsc{P@1} & \textsc{P@5} & \textsc{P@10} \\

\specialrule{1.2pt}{0.2em}{0.2em}

\multirow{5}{*}{Qw-3} & Base & 62.9\% & 75.0\% & 77.3\% & 7.0\% & 27.7\% & 43.3\% & 20.8\% & 31.9\% & 35.8\% & 1.8\% & 7.6\% & 12.7\% \\
 & Struct & \textbf{65.8\%} & 79.5\% & 82.1\% & 51.4\% & 72.4\% & 77.9\% & 21.3\% & 33.8\% & 39.1\% & \textbf{23.5\%} & 33.8\% & 37.7\% \\
 & Struct CoT & 64.2\% & 76.6\% & 79.3\% & 50.0\% & 72.4\% & 79.5\% & 20.8\% & 33.7\% & 38.9\% & 21.0\% & \textbf{34.5\%} & \textbf{39.5\%} \\
 & Refine & 63.7\% & \textbf{81.2\%} & \textbf{84.6\%} & 46.7\% & 68.1\% & 74.4\% & 19.0\% & 30.6\% & 34.8\% & 18.4\% & 29.9\% & 34.3\% \\
 & Refine CoT & 50.7\% & 71.5\% & 76.7\% & 42.7\% & 67.1\% & 74.6\% & 16.7\% & 28.9\% & 33.6\% & 15.4\% & 28.8\% & 34.7\% \\

\specialrule{1.2pt}{0.2em}{0.2em}

\multirow{5}{*}{Qw-7} & Base & 41.1\% & 72.2\% & 80.1\% & 16.0\% & 48.4\% & 63.5\% & \textbf{20.7\%} & 39.5\% & 46.3\% & 9.7\% & 29.7\% & 39.8\% \\
 & Struct & \textbf{48.8\%} & 77.5\% & 83.7\% & 48.1\% & 77.3\% & 84.2\% & 19.2\% & 35.8\% & 42.3\% & 19.2\% & 37.6\% & 45.6\% \\
 & Struct CoT & 43.9\% & 77.2\% & 84.8\% & 43.8\% & \textbf{78.4\%} & \textbf{86.0\%} & 20.5\% & \textbf{41.4\%} & \textbf{50.1\%} & 19.3\% & 39.0\% & 47.2\% \\
 & Refine & 41.6\% & 76.5\% & 83.8\% & 40.7\% & 75.9\% & 84.2\% & 19.7\% & 41.2\% & 49.7\% & 18.6\% & 39.5\% & 47.7\% \\
 & Refine CoT & 30.6\% & 69.5\% & 80.1\% & 30.4\% & 69.1\% & 79.6\% & 14.8\% & 36.2\% & 45.3\% & 17.0\% & 39.4\% & 49.1\% \\

\specialrule{1.2pt}{0.2em}{0.2em}

\multirow{5}{*}{Qw-14} & Base & 57.9\% & 86.9\% & 91.9\% & \textbf{69.1\%} & 89.0\% & 93.3\% & 49.4\% & 61.7\% & 64.7\% & \underline{\textbf{51.0\%}} & 62.5\% & 65.8\% \\
 & Struct & 53.6\% & 85.9\% & 91.9\% & 56.2\% & 87.7\% & 93.4\% & 43.5\% & 58.0\% & 61.9\% & 29.7\% & 56.3\% & 63.8\% \\
 & Struct CoT & 60.4\% & \textbf{89.1\%} & \textbf{94.3\%} & 60.0\% & 88.1\% & 93.0\% & 48.1\% & \underline{\textbf{65.3\%}} & \underline{\textbf{68.6\%}} & 32.7\% & 57.8\% & 65.4\% \\
 & Refine & 44.0\% & 83.8\% & 92.0\% & 43.7\% & 82.9\% & 90.1\% & 43.6\% & 61.5\% & 66.1\% & 27.0\% & 54.8\% & 63.2\% \\
 & Refine CoT & 42.8\% & 81.5\% & 89.3\% & 42.7\% & 82.2\% & 90.6\% & 44.0\% & 60.8\% & 66.2\% & 41.0\% & 57.8\% & 61.6\% \\

\specialrule{1.2pt}{0.2em}{0.2em}

\multirow{5}{*}{QwC-3} & Base & 42.8\% & 65.6\% & 72.5\% & 45.9\% & 71.7\% & 79.2\% & \textbf{18.8\%} & 33.5\% & 39.6\% & 14.2\% & 28.6\% & 36.0\% \\
 & Struct & \textbf{66.1\%} & \textbf{76.1\%} & 78.9\% & 48.7\% & 71.1\% & 77.9\% & 18.0\% & 32.5\% & 37.1\% & 17.8\% & 32.3\% & 36.7\% \\
 & Struct CoT & 43.9\% & 72.0\% & \textbf{79.9\%} & 39.7\% & 70.7\% & 78.9\% & 16.5\% & 33.3\% & 40.2\% & 17.1\% & \textbf{34.7\%} & \textbf{41.8\%} \\
 & Refine & 45.7\% & 68.7\% & 75.3\% & 46.4\% & 68.9\% & 75.1\% & 15.1\% & 30.1\% & 35.2\% & 16.0\% & 30.9\% & 35.8\% \\
 & Refine CoT & 40.0\% & 69.7\% & 77.0\% & 40.7\% & 70.0\% & 78.1\% & 13.0\% & 31.8\% & 39.7\% & 12.7\% & 31.1\% & 38.9\% \\

\specialrule{1.2pt}{0.2em}{0.2em}

\multirow{5}{*}{QwC-7} & Base & 61.5\% & 77.8\% & 82.0\% & 58.6\% & 77.2\% & 81.5\% & 31.4\% & 45.0\% & 48.1\% & 28.9\% & 46.0\% & 51.2\% \\
 & Struct & 64.9\% & 80.6\% & 84.3\% & \textbf{65.4\%} & 79.0\% & 82.2\% & 36.4\% & 44.5\% & 46.6\% & \textbf{36.9\%} & 44.8\% & 46.8\% \\
 & Struct CoT & 62.5\% & 83.4\% & 88.5\% & 64.4\% & 84.0\% & 89.0\% & 33.7\% & 49.8\% & 55.2\% & 32.5\% & 49.7\% & 54.6\% \\
 & Refine & 60.7\% & \textbf{84.5\%} & \textbf{89.6\%} & 61.3\% & 83.9\% & 88.5\% & 33.0\% & \textbf{52.6\%} & \textbf{58.5\%} & 32.1\% & 48.7\% & 54.6\% \\
 & Refine CoT & 49.3\% & 79.5\% & 85.3\% & 49.1\% & 79.0\% & 85.7\% & 26.2\% & 49.1\% & 56.3\% & 25.2\% & 49.5\% & 56.7\% \\

\specialrule{1.2pt}{0.2em}{0.2em}

\multirow{5}{*}{QwC-14} & Base & 71.8\% & 85.3\% & 88.5\% & 71.6\% & 86.0\% & 89.7\% & 36.1\% & 52.2\% & 55.8\% & 33.6\% & 53.4\% & 58.4\% \\
 & Struct & 75.1\% & 84.8\% & 87.1\% & \textbf{75.5\%} & 85.8\% & 87.7\% & \textbf{40.6\%} & 51.4\% & 53.1\% & 40.0\% & 52.3\% & 54.4\% \\
 & Struct CoT & 71.5\% & 87.6\% & \textbf{91.5\%} & 74.1\% & \textbf{88.5\%} & \textbf{91.5\%} & 37.8\% & 55.0\% & 60.1\% & 37.1\% & 52.1\% & 56.1\% \\
 & Refine & 69.2\% & 86.3\% & 89.8\% & 70.5\% & 86.7\% & 90.4\% & 38.0\% & 53.1\% & 57.7\% & 38.9\% & 53.8\% & 58.8\% \\
 & Refine CoT & 58.0\% & 83.1\% & 88.1\% & 59.2\% & 83.8\% & 88.6\% & 33.7\% & 53.0\% & 58.3\% & 35.6\% & \textbf{55.2\%} & \textbf{60.9\%} \\

\specialrule{1.2pt}{0.2em}{0.2em}

\multirow{5}{*}{DSC-16} & Base & 67.3\% & 82.8\% & 86.0\% & 71.8\% & 81.7\% & 84.4\% & 42.8\% & 53.9\% & 56.5\% & \textbf{44.9\%} & 53.7\% & 57.0\% \\
 & Struct & 24.9\% & 44.0\% & 52.0\% & 26.0\% & 44.9\% & 52.8\% & 13.8\% & 31.6\% & 39.6\% & 13.5\% & 30.3\% & 36.7\% \\
 & Struct CoT & \textbf{73.8\%} & \textbf{87.3\%} & \textbf{89.9\%} & 71.4\% & 87.1\% & \textbf{89.9\%} & 42.7\% & 55.7\% & 60.2\% & 43.1\% & \textbf{57.4\%} & \textbf{62.1\%} \\
 & Refine & 54.6\% & 78.7\% & 83.0\% & 55.0\% & 79.2\% & 83.2\% & 20.5\% & 41.5\% & 48.3\% & 21.1\% & 41.7\% & 48.0\% \\
 & Refine CoT & 59.0\% & 85.0\% & 89.5\% & 58.1\% & 85.0\% & 89.5\% & 34.0\% & 53.9\% & 59.4\% & 34.4\% & 53.9\% & 59.7\% \\

\specialrule{1.2pt}{0.2em}{0.2em}

\multirow{5}{*}{VR-3} & Base & 53.4\% & 73.6\% & 77.5\% & 61.2\% & 75.0\% & 79.6\% & 24.9\% & 42.3\% & 48.0\% & \textbf{31.0\%} & 45.1\% & 49.3\% \\
 & Struct & 58.5\% & 84.2\% & 89.2\% & 56.7\% & 83.9\% & 89.9\% & 29.0\% & 46.8\% & 52.4\% & 29.6\% & \textbf{49.3\%} & 55.7\% \\
 & Struct CoT & \textbf{65.1\%} & \textbf{87.5\%} & 92.3\% & 60.0\% & 87.0\% & \textbf{92.4\%} & 26.7\% & 49.1\% & \textbf{56.7\%} & 26.3\% & 46.1\% & 52.1\% \\
 & Refine & 24.2\% & 51.5\% & 62.4\% & 23.6\% & 55.7\% & 68.9\% & 10.8\% & 30.4\% & 40.1\% & 10.3\% & 31.1\% & 42.4\% \\
 & Refine CoT & 38.3\% & 72.7\% & 82.1\% & 31.0\% & 67.8\% & 80.0\% & 16.1\% & 37.1\% & 46.4\% & 17.3\% & 38.9\% & 49.2\% \\

\specialrule{1.2pt}{0.2em}{0.2em}

\multirow{5}{*}{VR-7} & Base & 59.1\% & 80.3\% & 86.7\% & 28.7\% & 66.1\% & 76.8\% & 34.4\% & 52.4\% & 57.8\% & 24.9\% & 46.4\% & 53.4\% \\
 & Struct & 66.2\% & 84.1\% & 88.9\% & \textbf{66.6\%} & 84.8\% & 89.6\% & 36.0\% & 52.1\% & 56.4\% & \textbf{37.5\%} & 53.0\% & 57.1\% \\
 & Struct CoT & 62.6\% & \textbf{87.9\%} & \textbf{92.9\%} & 63.2\% & 87.2\% & 91.6\% & 32.7\% & 53.5\% & 58.5\% & 31.8\% & \textbf{53.7\%} & \textbf{60.2\%} \\
 & Refine & 12.5\% & 35.6\% & 48.4\% & 12.3\% & 31.7\% & 42.7\% & 5.4\% & 19.8\% & 29.7\% & 3.9\% & 14.2\% & 21.2\% \\
 & Refine CoT & 49.7\% & 82.6\% & 89.5\% & 52.7\% & 83.4\% & 89.6\% & 29.5\% & 50.8\% & 56.2\% & 30.2\% & 51.5\% & 57.4\% \\

\specialrule{1.2pt}{0.2em}{0.2em}

\multirow{5}{*}{VT-7} & Base & \underline{\textbf{76.4\%}} & \underline{\textbf{94.1\%}} & \underline{\textbf{95.6\%}} & 75.3\% & 91.2\% & 93.5\% & 42.5\% & \textbf{63.5\%} & \textbf{66.5\%} & \textbf{44.3\%} & 59.5\% & 62.9\% \\
 & Struct & 43.3\% & 70.4\% & 79.0\% & 43.1\% & 70.1\% & 78.2\% & 21.5\% & 39.7\% & 45.2\% & 9.9\% & 28.5\% & 37.7\% \\
 & Struct CoT & 23.8\% & 55.2\% & 67.9\% & 24.1\% & 56.8\% & 69.4\% & 15.6\% & 37.0\% & 45.9\% & 16.1\% & 37.0\% & 45.8\% \\
 & Refine & 32.4\% & 61.6\% & 70.4\% & 31.7\% & 61.6\% & 70.1\% & 15.3\% & 37.5\% & 45.5\% & 17.2\% & 38.6\% & 46.3\% \\
 & Refine CoT & 14.6\% & 43.1\% & 56.3\% & 15.0\% & 44.6\% & 58.6\% & 8.1\% & 25.9\% & 35.1\% & 21.7\% & 40.6\% & 46.4\% \\

\specialrule{1.2pt}{0.2em}{0.2em}

\multirow{5}{*}{VT-14} & Base & \textbf{74.9\%} & \textbf{92.3\%} & \textbf{95.3\%} & 68.8\% & 90.3\% & 94.2\% & \textbf{38.1\%} & \textbf{55.9\%} & \textbf{60.0\%} & 26.2\% & 48.5\% & 55.9\% \\
 & Struct & 32.3\% & 59.1\% & 66.8\% & 23.6\% & 47.5\% & 56.5\% & 13.0\% & 23.3\% & 28.5\% & 13.4\% & 25.0\% & 30.5\% \\
 & Struct CoT & 17.9\% & 48.0\% & 60.6\% & 13.8\% & 40.5\% & 53.0\% & 10.1\% & 24.8\% & 32.9\% & 10.1\% & 26.5\% & 35.3\% \\
 & Refine & 29.5\% & 56.7\% & 66.0\% & 33.6\% & 61.1\% & 70.1\% & 17.2\% & 35.3\% & 42.9\% & 17.2\% & 35.3\% & 42.9\% \\
 & Refine CoT & 20.0\% & 50.8\% & 62.5\% & 19.6\% & 50.0\% & 61.5\% & 10.9\% & 31.3\% & 40.4\% & 9.7\% & 28.9\% & 38.8\% \\

\specialrule{1.2pt}{0.2em}{0.2em}

\end{tabular}
\vspace{-1em}
\end{table*}

\clearpage

\bibliographystyle{IEEEtran}
\bibliography{IEEEabrv,main}

\vfill

\end{document}